\documentclass[a4paper,12pt]{article}
\usepackage{graphicx}
\usepackage{float}
\usepackage{authblk}
\usepackage{etoolbox}
\usepackage[euler]{textgreek}
\usepackage{fixltx2e}
\usepackage{cite}
\usepackage[figuresonly,nolists,nomarkers]{endfloat}

\begin{document}

\title {Facile deterministic cutting of 2D materials for twistronics using a tapered fibre scalpel }

\author[1]{L.D. Varma Sangani}
\author[1]{R.S. Surya Kanthi}
\author[1]{Pratap Chandra Adak}
\author[1]{Subhajit Sinha}
\author[1]{Alisha H. Marchawala}
\author[2]{Takashi Taniguchi}
\author[2]{Kenji Watanabe}
\author[1]{Mandar M. Deshmukh\thanks{deshmukh@tifr.res.in}}
\affil[1]{Department of Condensed Matter Physics and Materials Science, Tata Institute of Fundamental Research, Homi Bhabha Road, Mumbai 400005, India.}
\affil[2]{National Institute for Materials Science (NIMS), Tsukuba, Ibaraki 305-0044, Japan.}

\maketitle

\section*{Abstract}
We present a quick and reliable method to cut 2D materials for creating 2D twisted heterostructures and devices. We demonstrate the effectiveness of using a tapered fibre scalpel for cutting graphene. Electrical transport measurements show evidence of the desired twist between the graphene layers fabricated using our technique. Statistics of the number of successfully twisted stacks made using our method is compared with h-BN assisted tear-and-stack method. Also, our method can be used for twisted stack fabrication of materials that are few nanometers thick. Finally, we demonstrate the versatility of the tapered fibre scalpel for other shaping related applications for sensitive 2D materials.


Two dimensional (2D) materials harbour a host of exciting phenomena \cite{doi:10.1146/annurev-matsci-070214-021034}. The newly discovered unconventional properties of these materials are of interest to multiple disciplines. There is an interest in the physics of excitons in transition metal dichalcogenides (TMDC) moiré superlattices \cite{tang2019wse2}. Recently, twisted bilayer graphene devices, in particular, have risen to prominence due to the presence of flat bands that can host correlated states.

 For the case of monolayer graphene based twistronic devices, a precise twist angle, called ‘magic angle', host a flat band that results in correlated insulating states \cite{Cao_2018}. Since the discovery of correlated insulating states in twisted graphene devices, further studies on twisted bilayer graphene have discovered signatures of superconductivity \cite{cao_unconventional_2018} and ferromagnetism \cite{sharpe_emergent_2019}. Theoretical proposals and initial experiments also suggest that introducing a twist angle in between other 2D heterostructure layers is a powerful knob to access flat band physics in other 2D materials like bilayer graphene \cite{chebrolu_flat_2019,burg_correlated_2019}, trilayer graphene \cite{tsai2019correlated} and TMDCs \cite{wang2019magic , tang2019wse2 , PhysRevLett.122.086402}.


While many different techniques to create 2D van der Waals heterostructures existed \cite{wang2015electronic, castellanos2014deterministic, dean2010boron, zomer2011transfer}, creating twisted heterostructures remained a challenge until recently \cite{kim_van_2016}.Introducing a deterministic twist angle between two layers requires knowledge of the relative orientation of their crystallographic axes. There is no easy way to get directions of crystallographic axes with sufficient accuracy reliably. A simple solution is to cut a large flake into two pieces  \cite{kim_van_2016} and place them on top of each other with a rotation angle introduced between them. This ensures that the crystallographic axes in the two sections are perfectly aligned before ‘twisting’.  Twisted 2D heterostructures are typically made using h-BN assisted tear-and-stack method \cite{kim_van_2016}, which is a challenging and probabilistic process.
An alternate way of controllably making twisted structures is to slice graphene before making the twisted heterostructure. Most exfoliated flakes are small, of the order of 10~\textmu m, so precise control is required to cut flakes. AFM cutting \cite{giesbers_nanolithography_2008} and electron beam lithography (EBL) patterning \cite{kim_van_2016} followed by etching of the flake are some methods to cut a flake. However, using AFM or EBL for sectioning may leave residue and is time-consuming. There is a need for a reliable and facile technique for cutting 2D flakes.


In this paper, we present a simple method to section 2D flakes using tapered fibre scalpel (TFS), which can be integrated with the microscope used for stacking. We demonstrate the sectioning of graphene flakes (dimensions of about 30~\textmu m) and make twisted double bilayer graphene devices. The TFS is very versatile, and we also use it to shape other sensitive 2D materials which might get contaminated by fabrication using lithographic techniques. Further, the TFS can also be used to cut materials a few nanometers thickness with ease.


\begin{figure}
\centering
\includegraphics[width=\linewidth]{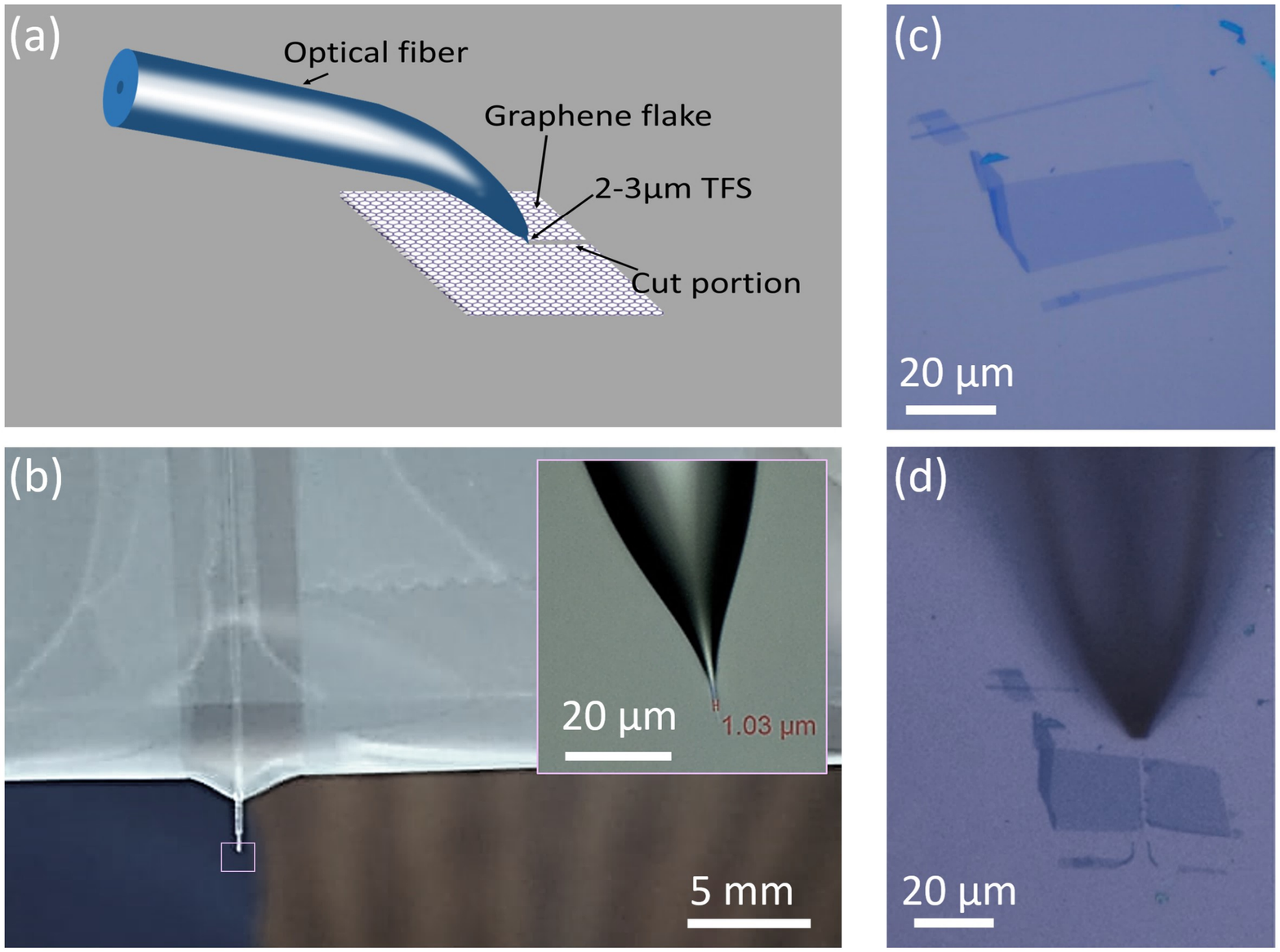}
\caption{Cutting flakes with TFS and integration with stacking microscope:(a) Schematic of the cutting process using a Tapered Fibre Scalpel (TFS).  (b) Photograph of the TFS mounted like a cantilever on a glass slide. The TFS is mounted on a glass slide and bent using scotch tape. The glass slide is then mounted on the micrometer stage to cut flakes. (inset): Optical image of the TFS. (c) and (d) Image of a graphene flake before cutting and after cutting with the TFS respectively.}
\label{fig2}
\end{figure}


Optical fibres can be tapered using a fibre pulling process that is used extensively for many applications \cite{Valaskovic:95}. We use these tapered optical fibres (details of optical fibre preparation are provided in S1 of supplementary information) as scalpels for slicing 2D materials. The schematic of the cutting process is depicted in Fig.~\ref{fig2}(a).

We exfoliate h-BN and graphene flakes on 280~nm SiO\textsubscript{2}/Si substrate. The TFS is placed on a glass slide like a cantilever, as shown in Fig.~\ref{fig2}(b), and the glass slide is fixed on the motorized micrometer precision stage, also used for stacking. We then bring the TFS near the flake and slowly raise the bottom stage containing the chip. Using the focus of the optical microscope, we determine when the chip on the bottom stage comes in contact with the TFS; after this, the flake is cut. We move the TFS in the direction we want to cut at a low speed of 1~\textmu m/s. Higher speed may lead to folding near the edge of the flake.  Fig.~\ref{fig2}(c) and Fig.~\ref{fig2}(d) are the optical images of the bilayer graphene flake before and after cutting using TFS.

By changing the relative orientation of the chip before cutting, and moving with low speed while cutting, we can slice graphene flakes along any direction. Further, this method of slicing of flakes makes use of a stacking stage, which is readily available for making van der Waals heterostructures. The TFS is easy to fabricate, and also ensures that the flake is minimally contaminated.

More details of the cutting procedure can be found in S1 of supplementary information.


\begin{figure}
\centering
\includegraphics[width=0.6\linewidth]{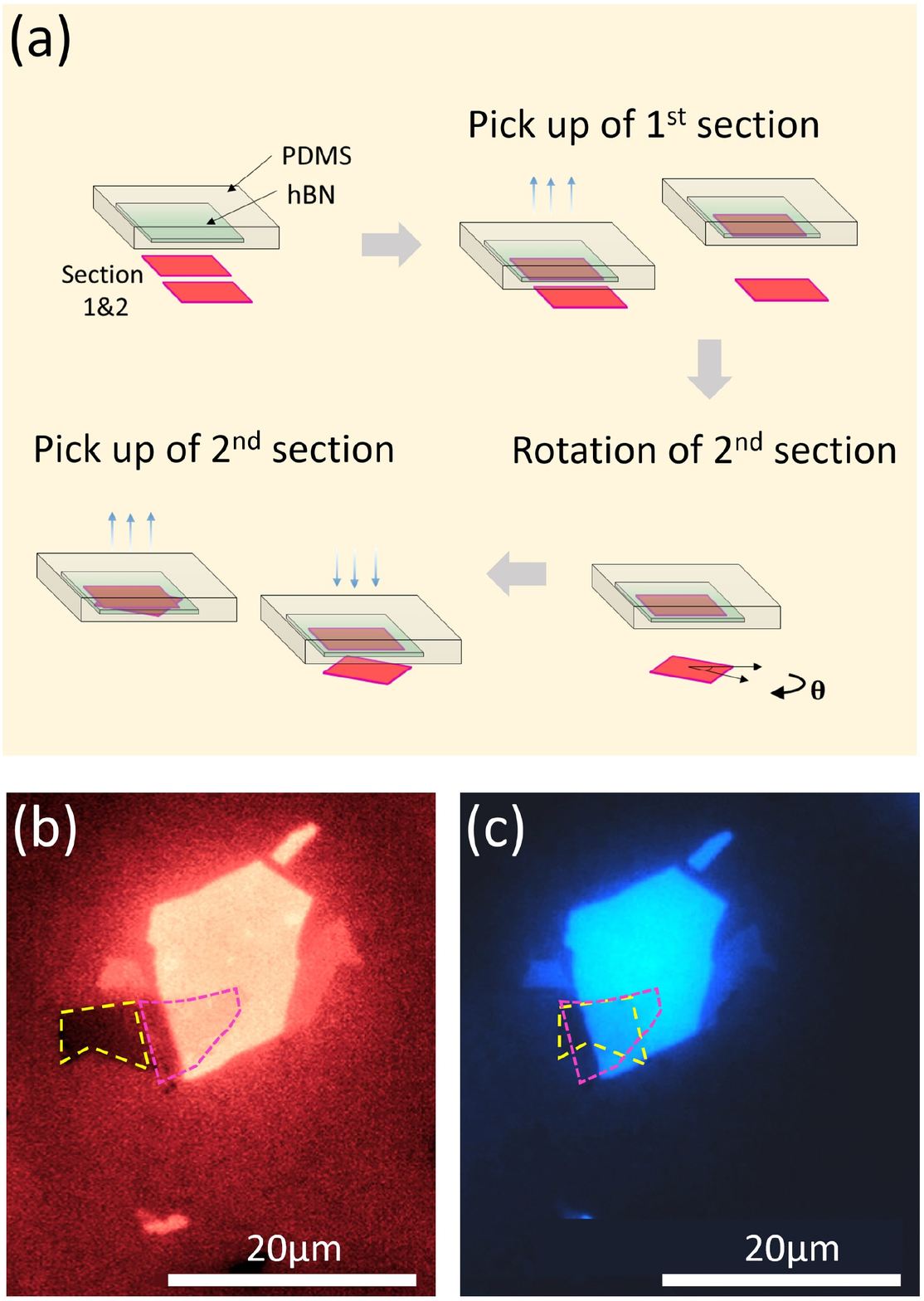}
\caption{Twisted heterostructure stacking using slice-and-stack: (a) Schematic of stacking with a sliced graphene flake for twisted graphene heterostructure. (b) Colour enhanced optical image of a sliced graphene flake being picked up by h-BN assisted van der Waals heterostructure stacking with the cut visible. Yellow and pink dashed lines indicate the boundaries of the sliced graphene. (c) Colour enhanced optical image of the second graphene transferred after rotation. Yellow and pink dashed lines indicate boundaries of the sliced graphene.}
\label{fig3}
\end{figure}


After slicing the graphene, the twisted devices are made using conventional van der Waals pick up technique. Fig.~\ref{fig3}(a) shows a schematic of the pick-up process of the two portions of the cut graphene flake. To prepare the twisted graphene stack, we first align the h-BN on a PC-PDMS stamp with the graphene such that it covers only one section of the sliced graphene flake \cite{kim_van_2016} as shown in Fig.~\ref{fig3}(b). We thus pick up one cut portion of the graphene. We then rotate the chip with the second portion of graphene by a specified angle $\theta$~and pick-up the second half of the graphene using van der Waals forces to complete the twisted stack as shown in Fig.~\ref{fig3}(c). The color enhanced optical micrographs from Fig.~\ref{fig3}(b) and Fig.~\ref{fig3}(c) can be found in S4 of the supplementary information.
We make Hall bar devices of the twisted double bilayer stacks and electrically characterize them. The twisted devices made show moiré bands, proving that it is indeed twisted. We present evidence of the moiré peaks in one of the devices made using the slice-and-stack method subsequently.


\begin{table*}[t]
  \begin{center}
    \caption{Comparison table of success of first half flake pickup using tear-and-stack and using slice-and-stack method.}
    \label{table1}
    \begin{tabular}{|c|c|c|} 
    \hline
      \textbf{Method} & \textbf{Twisted stack number} & \textbf{1\textsuperscript{st} section picked up at}\\
      \hline
      Tear-and-stack & Stack 1 & 6\textsuperscript{th} attempt \\
       & Stack 2 & 3\textsuperscript{rd} attempt \\
       & Stack 3 & 7\textsuperscript{th} attempt \\
       & Stack 4-10 & Not successful \\
      \hline
      TFS assisted slice-and-stack & Stack 1 & Not successful \\
      & Stack 2-10 & 1\textsuperscript{st} attempt \\
      \hline
      \end{tabular}
  \end{center}
\end{table*}

The statistics of 10 consecutive attempts of making twisted devices using the tear-and-stack method and slice-and-stack method is shown in Table~\ref{table1}. As is clear from the data, using the slice-and-stack method, we are able to make twisted stacks with ease in the first attempt reliably. The yield of the devices ‘twisted’ successfully is better than the tear-and-stack method. Additionally, unintentional straining in the stack \cite{saito2019decoupling} can result from the tear-and-stack method, which can be avoided using the slice-and-stack method.


Commonly used methods for cutting graphene before stacking includes AFM cutting \cite{ giesbers_nanolithography_2008} and EBL \cite{ kim_van_2016}. Cutting graphene with AFM takes a large amount of time (around an hour). EBL patterning is another reliable method to slice flakes, but is time-consuming and can cause contamination of flakes. Using the TFS, we can achieve rapid cutting of graphene flakes (at a speed of 1~\textmu m/s) with minimal contamination.


\begin{figure}
\centering
\includegraphics[width=\linewidth]{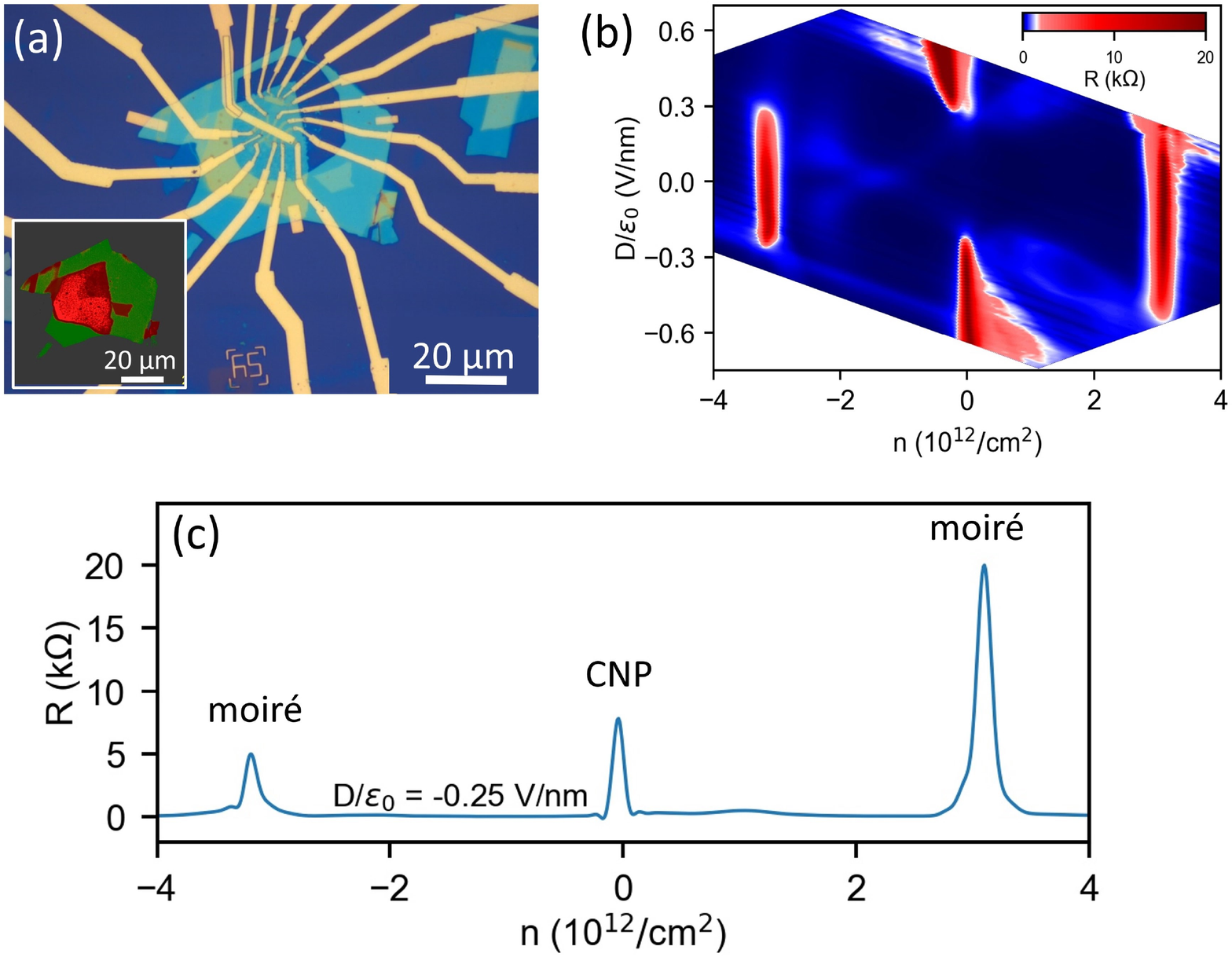}
\caption{Data from a twisted graphene device made using slice and stack method: (a) Optical micrograph of a fully fabricated dual gated twisted double bilayer graphene device made using the slice-and-stack technique. Dual gates allow independent control over the perpendicular electric field applied on the device and charge density of carriers in the device. (inset) Colour enhanced optical image of the twisted double bilayer graphene stack made using this method. The two graphene layers, covered by hBN from top and bottom sides, are clearly visible in the tweaked image. (b) A color scale plot of 4-probe resistance of the device having a twist angle $\theta$ = 1.18$^\circ$, as a function of charge density (n) and perpendicular electric displacement field (D). It demonstrates features which are characteristic to a small-angle twisted double bilayer graphene device. (c) A line slice plot of resistance vs. charge density is shown for D/\textepsilon\textsubscript{0} = -0.25 V/nm, showing all the three resistance peaks corresponding to the charge neutrality point gap at n=0 and the two moiré gaps at n= $\pm$3.2×10\textsuperscript{12}~cm\textsuperscript{-2}.}

\label{fig4}
\end{figure}

Now, we discuss the electrical response of the fabricated twisted double-bilayer graphene device that proves the efficacy of our technique. Fig.~\ref{fig4}(a) shows image of a device fabricated using the TFS assisted slice-and-stack method. Fig.~\ref{fig4}(b) shows the 4-probe resistance as a function of gate induced charge density and perpendicular electric displacement field (D) of a twisted double bilayer graphene device fabricated using the TFS assisted slice-and-stack. The data from the twisted double bilayer graphene device can be used to extract the bandgaps and bandwidths of the system\cite{adak2020tunable}.

Fig.~\ref{fig4}(c) shows the device resistance as a function of gate voltage for a device made using this method. The moiré peaks are distinctly visible. Moiré patterns were consistently obtained in devices twisted with this method. More resistance as a function of gate voltage curves showing moiré peaks for other devices are provided in S3 of supplementary information.

\begin{figure}
\centering
\includegraphics[width=\linewidth]{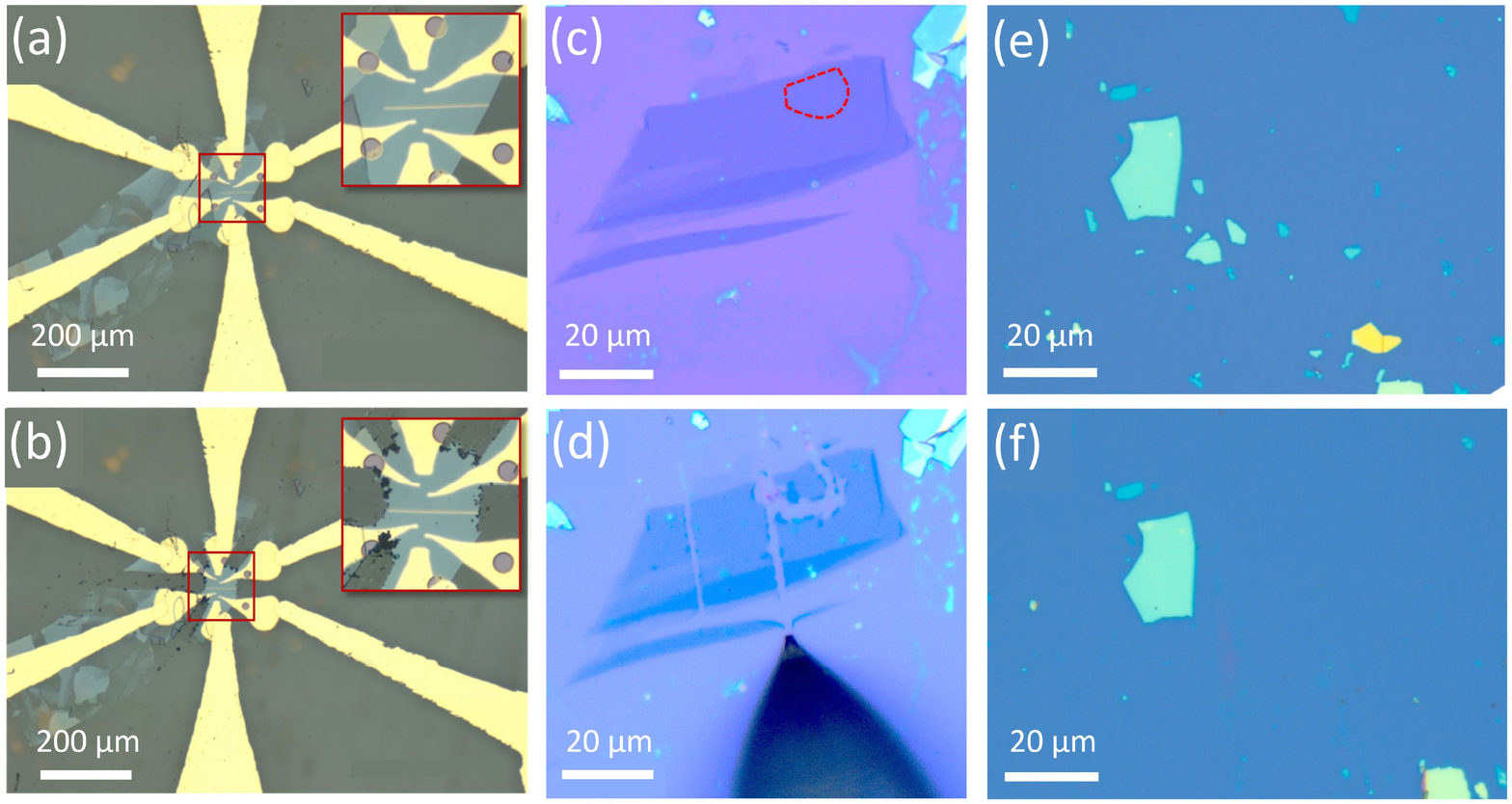}
\caption{Shaping other materials using TFS: (a) and (b) Optical images of the BSCCO device before and after shaping of BSCCO flake respectively. Inset: higher magnification image. (c) and (d) optical images of ABA and ABC flake before and after separating ABA and ABC portions respectively. Encircled region in (c) is ABC part which was identified using Raman analysis. (e) and (f) optical images of h-BN flake before and after cleaning its surroundings respectively.}

\label{fig5}
\end{figure}

Thus far, we have used the TFS for slicing graphene quite frequently. However, the TFS can also be used for shaping other 2D flakes in a regular desired geometry. For example, 2D materials like the high T\textsubscript{c} superconductor~Bi\textsubscript{2}Sr\textsubscript{2}Ca\textsubscript{2}Cu\textsubscript{3}O\textsubscript{10+\textdelta} (BSCCO) in a few unit-cell limit are of great interest \cite{zhao_sign-reversing_2019,yu_high-temperature_2019,doi:10.1021/acs.nanolett.8b02183}. However, the time-dependent degradation of BSCCO in ambient conditions\cite{zhao_sign-reversing_2019} makes it difficult to work with. The minimal invasiveness and rapidity of cutting with the TFS, along with its simple integration with the micrometer precision stage used for stacking, make the TFS ideal for shaping purposes. We regularly use the TFS to rapidly shape BSCCO flakes without using any chemicals. This shaping step is crucial to our devices, as it helps us control the device geometry. Fig.~\ref{fig5}(a) and Fig.~\ref{fig5}(b) show an optical micrograph of a BSCCO device before and after shaping respectively.


The TFS were also used in isolation of an area with ABC stacked trilayer graphene stacking in a larger flake. ABC trilayer is typically a smaller fraction of exfoliated flake and easily converts to ABA graphene on heating \cite{chen_signatures_2019} if in contact with ABA trilayer graphene.  So, it is necessary for it to be isolated from surrounding ABA graphene regions before being fabricated into a device. Traditional lithography techniques involve heating and chemicals, which can result in contamination. With the TFS, we easily isolate the ABC trilayer graphene from the ABA trilayer graphene. Fig.~\ref{fig5}(c) and Fig.~\ref{fig5}(d) show the trilayer graphene flake before and after the ABC trilayer region is separated out.

In general, stacking picks up the flakes in the neighborhood of our flake of interest, which can be an issue while making devices. We remove any unwanted flakes around our flake of interest easily with the help of the TFS and a micrometer precision stage. Fig.~\ref{fig5}(e) and Fig.~\ref{fig5}(f) show images of a h-BN flake before and after its surroundings are cleaned with the TFS respectively. Other examples of slicing 2D flakes are provided in S2 of supplementary information.

We have demonstrated the efficacy of the slice-and-stack technique of making twisted graphene stacks. Additionally, the slice-and-stack technique can be used for graphene slicing with a TFS for rapidity, reliability, and minimal contamination.
The tapered fibre scalpel described is easily integrated with existing stacking setups and easy to use. In general, the TFS can be used for slicing any 2D materials rapidly and can also be used to create twisted stacks of other few nanometer-thick materials like transition metal dichalcogenides (TMDCs). Additionally, it can also be used for cleaning areas near a flake of interest while stacking.

\section*{Acknowledgement}
We thank Sanat Ghosh, Jhuma Saha for helpful discussion and experimental assistance.  We
acknowledge the Swarnajayanti Fellowship of the Department of Science
and Technology (for M.M.D.), DST Nanomission grant SR/NM/NS-45/2016,
and the Department of Atomic Energy of the Government of India for
support.


\begin{thebibliography}{10}

\bibitem{doi:10.1146/annurev-matsci-070214-021034}
S.~Das, J.~A. Robinson, M.~Dubey, H.~Terrones, and M.~Terrones, ``Beyond
  graphene: Progress in novel two-dimensional materials and van der waals
  solids,'' {\em Annual Review of Materials Research}, vol.~45, no.~1,
  pp.~1--27, 2015.

\bibitem{tang2019wse2}
Y.~Tang, L.~Li, T.~Li, Y.~Xu, S.~Liu, K.~Barmak, K.~Watanabe, T.~Taniguchi,
  A.~H. MacDonald, J.~Shan, {\em et~al.}, ``Wse2/ws2 moir\'e superlattices: a
  new hubbard model simulator.'' 2019.

\bibitem{Cao_2018}
Y.~Cao, V.~Fatemi, A.~Demir, S.~Fang, S.~L. Tomarken, J.~Y. Luo, J.~D.
  Sanchez-Yamagishi, K.~Watanabe, T.~Taniguchi, E.~Kaxiras, and et~al.,
  ``Correlated insulator behaviour at half-filling in magic-angle graphene
  superlattices,'' {\em Nature}, vol.~556, p.~80–84, Mar 2018.

\bibitem{cao_unconventional_2018}
Y.~Cao, V.~Fatemi, S.~Fang, K.~Watanabe, T.~Taniguchi, E.~Kaxiras, and
  P.~Jarillo-Herrero, ``Unconventional superconductivity in magic-angle
  graphene superlattices,'' {\em Nature}, vol.~556, pp.~43--50, Apr. 2018.

\bibitem{sharpe_emergent_2019}
A.~L. Sharpe, E.~J. Fox, A.~W. Barnard, J.~Finney, K.~Watanabe, T.~Taniguchi,
  M.~A. Kastner, and D.~Goldhaber-Gordon, ``Emergent ferromagnetism near
  three-quarters filling in twisted bilayer graphene,'' {\em Science},
  vol.~365, pp.~605--608, Aug. 2019.

\bibitem{chebrolu_flat_2019}
N.~R. Chebrolu, B.~L. Chittari, and J.~Jung, ``Flat bands in twisted double
  bilayer graphene,'' {\em Physical Review B}, vol.~99, p.~235417, June 2019.

\bibitem{burg_correlated_2019}
G.~W. Burg, J.~Zhu, T.~Taniguchi, K.~Watanabe, A.~H. MacDonald, and E.~Tutuc,
  ``Correlated {Insulating} {States} in {Twisted} {Double} {Bilayer}
  {Graphene},'' {\em Physical Review Letters}, vol.~123, p.~197702, Nov. 2019.

\bibitem{tsai2019correlated}
K.-T. Tsai, X.~Zhang, Z.~Zhu, Y.~Luo, S.~Carr, M.~Luskin, E.~Kaxiras, and
  K.~Wang, ``Correlated superconducting and insulating states in twisted
  trilayer graphene moire of moire superlattices.'' 2019.

\bibitem{wang2019magic}
L.~Wang, E.-M. Shih, A.~Ghiotto, L.~Xian, D.~A. Rhodes, C.~Tan, M.~Claassen,
  D.~M. Kennes, Y.~Bai, B.~Kim, {\em et~al.}, ``Magic continuum in twisted
  bilayer wse2.'' 2019.

\bibitem{PhysRevLett.122.086402}
F.~Wu, T.~Lovorn, E.~Tutuc, I.~Martin, and A.~H. MacDonald, ``Topological
  insulators in twisted transition metal dichalcogenide homobilayers,'' {\em
  Phys. Rev. Lett.}, vol.~122, p.~086402, Feb 2019.

\bibitem{wang2015electronic}
J.~I.-J. Wang, Y.~Yang, Y.-A. Chen, K.~Watanabe, T.~Taniguchi, H.~O. Churchill,
  and P.~Jarillo-Herrero, ``Electronic transport of encapsulated graphene and
  wse2 devices fabricated by pick-up of prepatterned hbn,'' {\em Nano letters},
  vol.~15, no.~3, pp.~1898--1903, 2015.

\bibitem{castellanos2014deterministic}
A.~Castellanos-Gomez, M.~Buscema, R.~Molenaar, V.~Singh, L.~Janssen, H.~S. Van
  Der~Zant, and G.~A. Steele, ``Deterministic transfer of two-dimensional
  materials by all-dry viscoelastic stamping,'' {\em 2D Materials}, vol.~1,
  no.~1, p.~011002, 2014.

\bibitem{dean2010boron}
C.~R. Dean, A.~F. Young, I.~Meric, C.~Lee, L.~Wang, S.~Sorgenfrei, K.~Watanabe,
  T.~Taniguchi, P.~Kim, K.~L. Shepard, {\em et~al.}, ``Boron nitride substrates
  for high-quality graphene electronics,'' {\em Nature nanotechnology}, vol.~5,
  no.~10, pp.~722--726, 2010.

\bibitem{zomer2011transfer}
P.~Zomer, S.~Dash, N.~Tombros, and B.~Van~Wees, ``A transfer technique for high
  mobility graphene devices on commercially available hexagonal boron
  nitride,'' {\em Applied Physics Letters}, vol.~99, no.~23, p.~232104, 2011.

\bibitem{kim_van_2016}
K.~Kim, M.~Yankowitz, B.~Fallahazad, S.~Kang, H.~C.~P. Movva, S.~Huang,
  S.~Larentis, C.~M. Corbet, T.~Taniguchi, K.~Watanabe, S.~K. Banerjee, B.~J.
  LeRoy, and E.~Tutuc, ``van der {Waals} {Heterostructures} with {High}
  {Accuracy} {Rotational} {Alignment},'' {\em Nano Letters}, vol.~16,
  pp.~1989--1995, Mar. 2016.

\bibitem{giesbers_nanolithography_2008}
A.~J.~M. Giesbers, U.~Zeitler, S.~Neubeck, F.~Freitag, K.~S. Novoselov, and
  J.~C. Maan, ``Nanolithography and manipulation of graphene using an atomic
  force microscope,'' {\em Solid State Communications}, vol.~147, pp.~366--369,
  Sept. 2008.

\bibitem{Valaskovic:95}
G.~A. Valaskovic, M.~Holton, and G.~H. Morrison, ``Parameter control,
  characterization, and optimization in the fabrication of optical fiber
  near-field probes,'' {\em Appl. Opt.}, vol.~34, pp.~1215--1228, Mar 1995.

\bibitem{saito2019decoupling}
Y.~Saito, J.~Ge, K.~Watanabe, T.~Taniguchi, and A.~F. Young, ``Decoupling
  superconductivity and correlated insulators in twisted bilayer graphene.''
  2019.

\bibitem{adak2020tunable}
P.~C. Adak, S.~Sinha, U.~Ghorai, L.~Sangani, K.~Watanabe, T.~Taniguchi,
  R.~Sensarma, and M.~M. Deshmukh, ``Tunable bandwidths and gaps in twisted
  double bilayer graphene system on the verge of correlations.'' 2020.

\bibitem{zhao_sign-reversing_2019}
S.~F. Zhao, N.~Poccia, M.~G. Panetta, C.~Yu, J.~W. Johnson, H.~Yoo, R.~Zhong,
  G.~Gu, K.~Watanabe, T.~Taniguchi, S.~V. Postolova, V.~M. Vinokur, and P.~Kim,
  ``Sign-{Reversing} {Hall} {Effect} in {Atomically} {Thin}
  {High}-{Temperature} { $Bi_{2.1}Sr_{1.9}CaCu_{2.0}O_{8+\delta}$ }
  {Superconductors},'' {\em Physical Review Letters}, vol.~122, p.~247001, June
  2019.

\bibitem{yu_high-temperature_2019}
Y.~Yu, L.~Ma, P.~Cai, R.~Zhong, C.~Ye, J.~Shen, G.~D. Gu, X.~H. Chen, and
  Y.~Zhang, ``High-temperature superconductivity in monolayer
  ${Bi_{2}Sr_2CaCu_2O_{8+\delta}}$,'' {\em Nature}, vol.~575, pp.~156--163,
  Nov. 2019.

\bibitem{doi:10.1021/acs.nanolett.8b02183}
M.~Liao, Y.~Zhu, J.~Zhang, R.~Zhong, J.~Schneeloch, G.~Gu, K.~Jiang, D.~Zhang,
  X.~Ma, and Q.-K. Xue, ``{Superconductor}–{Insulator} {Transitions} in
  {Exfoliated} {$Bi_2Sr_2CaCu_2O_{8+\delta}$} {Flakes},'' {\em Nano Letters},
  vol.~18, no.~9, pp.~5660--5665, 2018.

\bibitem{chen_signatures_2019}
G.~Chen, L.~Jiang, S.~Wu, B.~Lyu, H.~Li, B.~L. Chittari, K.~Watanabe,
  T.~Taniguchi, Z.~Shi, J.~Jung, Y.~Zhang, and F.~Wang, ``Evidence of a
  gate-tunable {Mott} insulator in a trilayer graphene moiré superlattice,''
  {\em Nature Physics}, vol.~15, pp.~237--241, Mar. 2019.

\end{thebibliography}

\end{document}


\section*{Supplementary Information}

\beginsupplement
\section{Making of tips}
\begin{figure}[H]
\centering
\includegraphics[width=\linewidth]{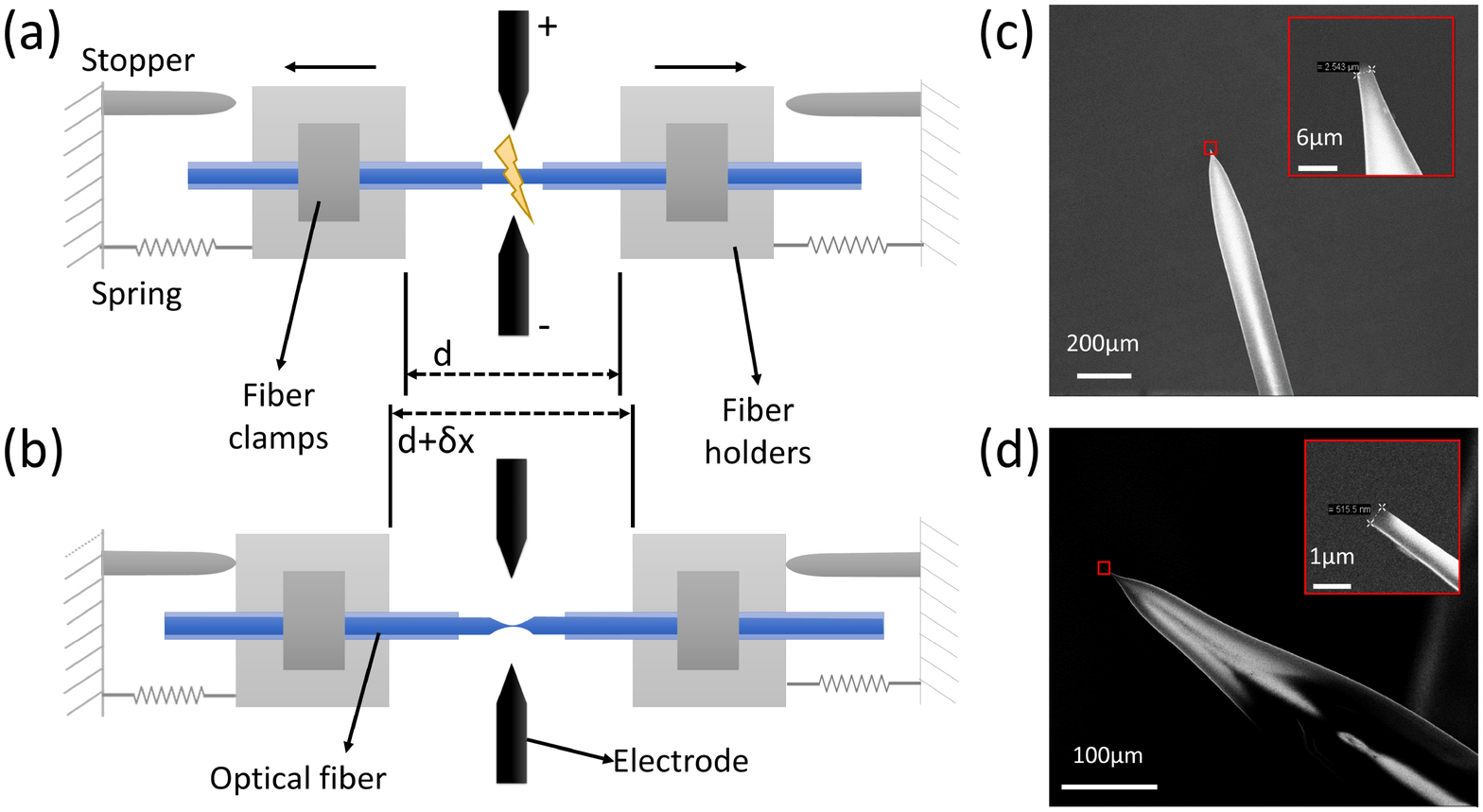}
\caption{Fibre tapering setup details: Schematic of the fibre tapering setup showing (a) position of clamps while arcing (b) position of clamps after arcing. (c) and (d) Image of a tapered fibre tip with a zoomed in image of the tip (inset).}
\label{figS1}
\end{figure}


We use a Fujikura 80S optical fibre splicer to taper optical fibres (Thorlabs SM980) and make sub-micrometer tapers for the TFS. Making the TFS involves 3 steps: building the tension, coarse tapering, and fine tapering of the fibre. We first remove the polymer coating on the fibre at a desired location and mount it on the movable holders of splicer as shown in Fig.~\ref{figS1}(a). The holders are spring loaded. Before loading, we move the holder by distance \textdelta x towards the centre using micrometre stage and hold it there. We load a fibre on the holders and clamp it on both sides. The holders are then released. This process creates some tension in the fibre and allows the holders to move away from centre on application of an electric arc which results in tapering. The schematic of the splicer setup before and after tapering is shown in Fig.~\ref{figS1}(a) and Fig.~\ref{figS1}(b) respectively. Tension developed in the fibre determines the final diameter of the tapering. We have achieved 78$\pm$4~\textmu m diameter for 0.5 mm movement and 36$\pm$6~\textmu m for 1 mm movement for single arc at STD-30 power for 300 ms arcing time.\\


While tapering the fibres, we have observed that greater tension leads to higher tapering and thinner diameters but tension if too high leads to fibre breakage and round edge formation during arcing, forming TFS with diameters more than 5~\textmu m \cite{Valaskovic:95}\cite{marchi_cartilage_2018}. Thus, we needed careful control of tension and arc power to successfully create few micron or even sub-micron TFS.
We use two step arcing to make the submicron TFS. This process results in formation of small diameter TFS reproducibly. We taper the fibre to a diameter of about 10~\textmu m by giving an optimized tension, and an electric arc with high arc time. We then apply several low arc time electric arcs which narrows the tip down to a few micron length scale. The 2 step process of tapering the fibre ensures that the length of the low diameter portion is small enough to make the tip effective yet long lasting and robust. The average minimum diameter we obtain is 1.5$\pm$1~\textmu m.\\


 SEM images of two TFS with different diameters 2.5~\textmu m and 0.5~\textmu m made using this process are shown in Fig.~\ref{figS1}(c) and Fig.~\ref{figS1}(d). We have achieved TFS with diameters as small as 500~nm using this two-step process. Nonetheless, we use TFS with 1 or 2~\textmu m for sectioning the flake because the visibility of the resulting cut portion is helpful for stacking and because we find them robust.

 \begin{figure}[H]
\centering
\includegraphics[width=\linewidth]{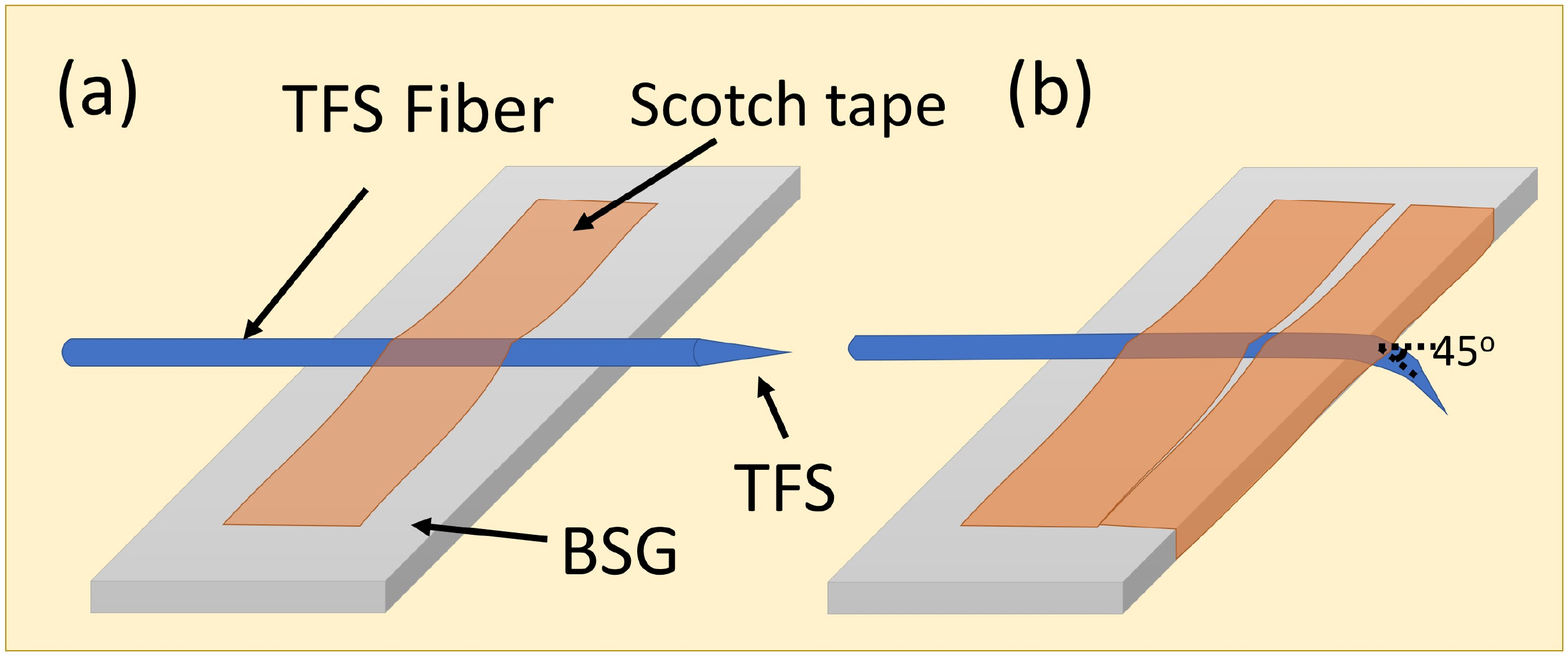}
\caption{Mounting TFS on a glass slide: (a) Schematic of the TFS on a glass slide before bending. (b) Schematic of the fully mounted TFS, bent with scotch tape.}
\label{fig_extra}
\end{figure}

The TFS is mounted on a glass slide using Scotch tape before cutting. First, we take a TFS fiber and fix it with scotch tape on a 75x25 mm\textsuperscript{2} borosilicate glass (BSG) slide. The TFS is fixed such that about 5~mm of the pointed end of the TFS is protruding from the longer edge of the BSG as shown in Fig.~\ref{fig_extra}(a). We then bend the hanging portion of the TFS using scotch tape till the angle between glass surface axis and hanging part is around 45$^\circ$, as shown in Fig.~\ref{fig_extra}(b). The edge of the scotch tape should be 2-3 mm away from the scalpel.


\section{Other examples of cut flakes}
\begin{figure}[H]
\centering
\includegraphics[width=0.8\linewidth]{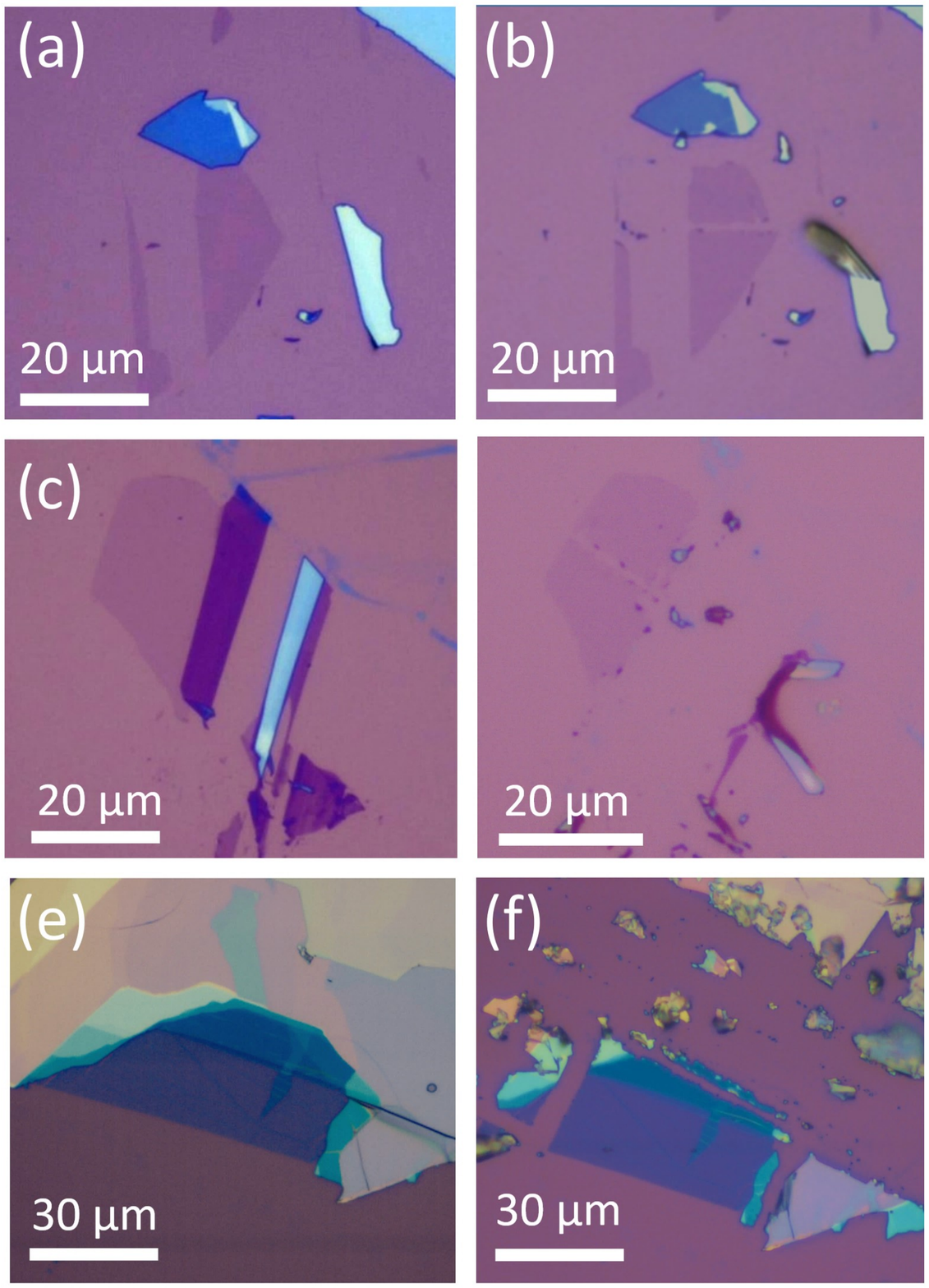}
\caption{Other flakes cut with TFS:(a) and (b) Graphene flakes before cutting with optical fibre tip. (c) and (d) Graphene flakes after cutting with optical fibre tip. The TFS can also be used for separating a thin graphene flake from a thick graphene flake, as demonstrated in (a) and (c). (e) and (f) Image of an MoS\textsubscript{2} flake before cutting and after cutting with TFS respectively.}
\label{figS2}
\end{figure}

\section{Data from some other twisted devices fabricated using slice-and-stack:}
\begin{figure}[H]
\centering
\includegraphics[width=\linewidth]{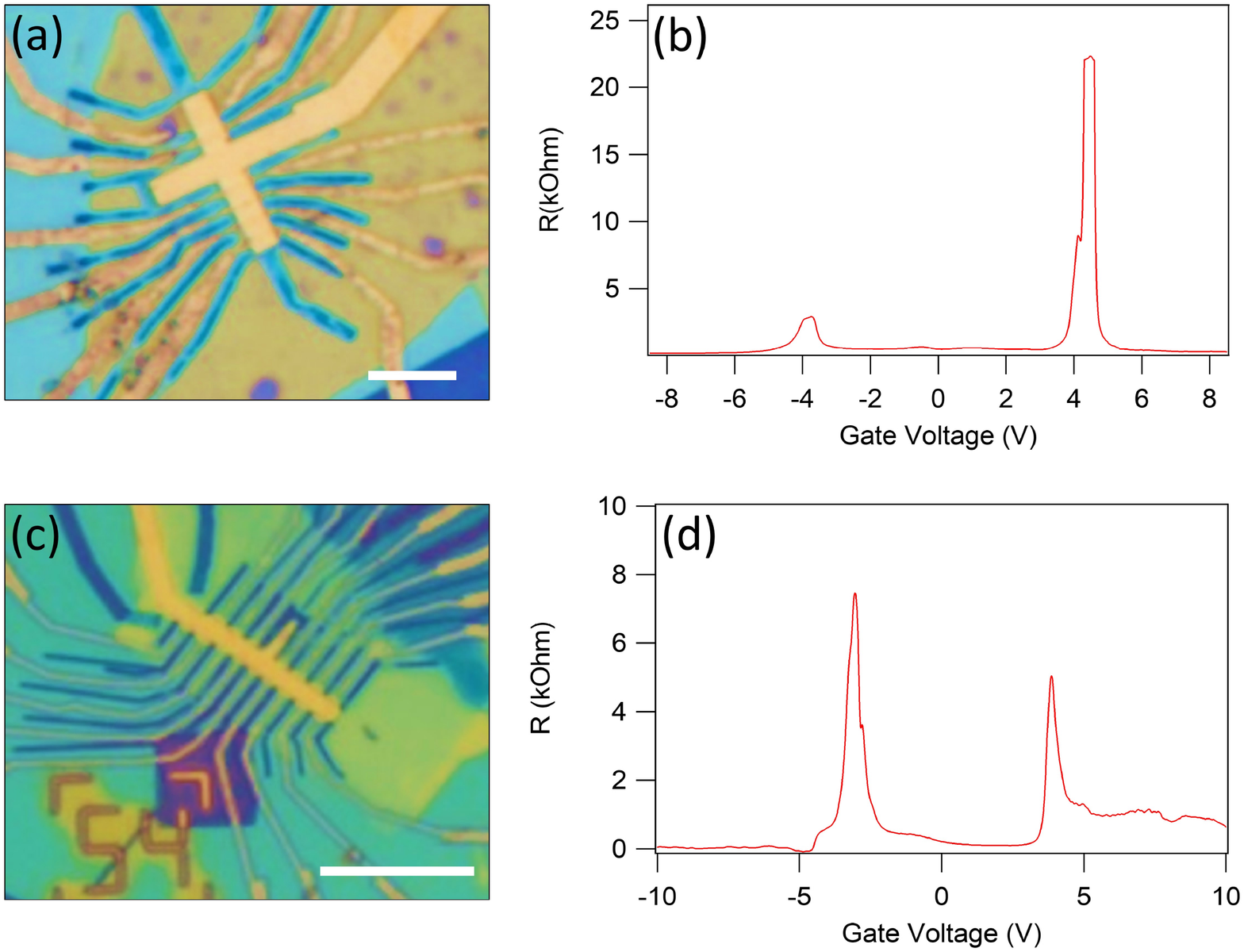}
\caption{Electrical measurements from some devices made using TFS assisted slice-and-stack: (a) Image of the twisted double bilayer graphene device 2 which is measured. Scale bar represents 10~\textmu m. (b) Resistance as a function of gate voltage for twisted double bilayer graphene device 2 fabricated using TFS assisted slice-and-stack. (c) Image of the twisted double bilayer graphene device 3 measured. Scale bar represents 10~\textmu m. (d) Resistance as a function of gate voltage for twisted double bilayer graphene device 3 fabricated using TFS assisted slice-and-stack.}
\label{figS3}
\end{figure}

\section{Miscellaneous}
\begin{figure}[H]
\centering
\includegraphics[width=\linewidth]{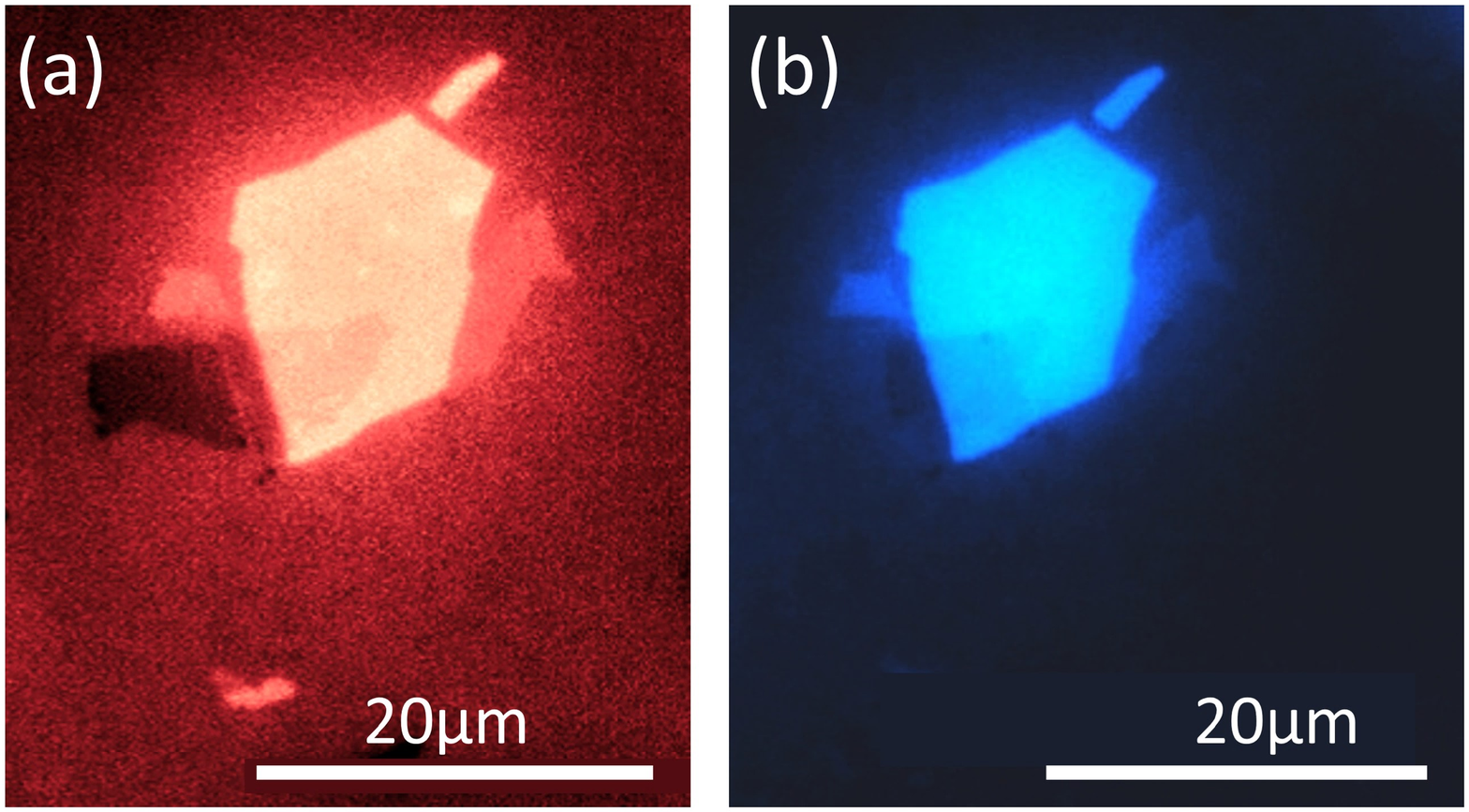}
\caption{Images from Fig 2 in main text without optical guides:(a) Colour enhanced image of the BN picking up first slice of graphene shown in Fig.~2(b) of the main text with the dashed lines omitted. (b)  Colour enhanced image of the BN picking up second slice of graphene to complete the twisted stack shown in Fig.~2(c) of the main text with the dashed lines omitted.}
\label{figS4}
\end{figure}